\documentclass{article}

\usepackage{arxiv}
\usepackage{eurosym}
\usepackage[utf8]{inputenc} 
\usepackage[T1]{fontenc}    
\usepackage{hyperref}       
\usepackage{url}            
\usepackage{booktabs}       
\usepackage{amsfonts}       
\usepackage{nicefrac}       
\usepackage{microtype}      
\usepackage{lipsum}
\usepackage{graphicx}
\usepackage{color}
\usepackage{amssymb}
\usepackage[ruled,vlined]{algorithm2e}
\usepackage{subcaption}
\usepackage{amsmath}
\usepackage[autostyle]{csquotes}

\title{Building power consumption datasets: Survey, taxonomy and future directions}

\author{
  Yassine Himeur\thanks{Energy and Buildings, vol. 227, 2020, 110404} , Abdullah Alsalemi, Faycal Bensaali\\
  Department of Electrical Engineering\\
  Qatar University\\
  Doha, Qatar \\
  \texttt{yassine.himeur@qu.edu.qa;a.alsalemi@qu.edu.qa;f.bensaali@qu.edu.qa} \\
   \And
 Abbes Amira \\
  Institute of Artificial Intelligence\\
  De Montfort University\\
  Leicester, United Kingdom \\
  \texttt{abbes.amira@dmu.ac.uk} \\
}

\begin{document}
\maketitle

\begin{abstract}
In the last decade, extended efforts have been poured into energy efficiency. Several energy consumption datasets were henceforth published, with each dataset varying in properties, uses and limitations. For instance, building energy consumption patterns are sourced from several sources, including ambient conditions, user occupancy, weather conditions and consumer preferences. Thus, a proper understanding of the available datasets will result in a strong basis for improving energy efficiency. Starting from the necessity of a comprehensive review of existing databases, this work is proposed to survey, study and visualize the numerical and methodological nature of building energy consumption datasets. A total of thirty-one databases are examined and compared in terms of several features, such as the geographical location, period of collection, number of monitored households, sampling rate of collected data, number of sub-metered appliances, extracted features and release date. Furthermore, data collection platforms and related modules for data transmission, data storage and privacy concerns used in different datasets are also analyzed and compared. Based on the analytical study, a novel dataset has been presented, namely Qatar university dataset, which is an annotated power consumption anomaly detection dataset. The latter will be very useful for testing and training anomaly detection algorithms, and hence reducing wasted energy. Moving forward, a set of recommendations is derived to improve datasets collection, such as the adoption of multi-modal data collection, smart Internet of things data collection, low-cost hardware platforms and privacy and security mechanisms. In addition, future directions to improve datasets exploitation and utilization are identified, including the use of novel machine learning solutions, innovative visualization tools and explainable mobile recommender systems. Accordingly, a novel visualization strategy based on using power consumption micro-moments has been presented along with an example of deploying machine learning algorithms to classify the micro-moment classes and identify anomalous power usage. 
\end{abstract}


\section{Introduction} \label{sec1}
Recent studies have shown that buildings are in charge of more than 40\% of power consumption demand and greenhouse gas emissions around the world. Indeed, this steady growth in energy consumption has been closely tied to the increasing number of population and the rising levels of prosperity \cite{CHAUDHURI2019,HIMEUR2020114877}. Moreover, even climate conditions in certain regions of the world obliged households to call for more energy for heating, cooling, cooking and refrigeration needs. Business buildings, principally offices and university structures, are also viewed as structures exhibiting high power consumption \cite{GUL2015,Zhang2018}. Consequently, the expansion of power usage and carbon emission as well as expensive energy prices in the above mentioned environments has made energy preserving a vital goal for various public authorities of all governments in order to accomplish efficient energy reduction \cite{BROGGER2018a,BROGGER2018b,MOHAMED2019,SHER2019,JAFARINEJAD2019}.

The building energy sector has recently attract the attention of various public energy efficiency initiatives to accomplish greater energy sustainability. Additionally, a clear correlation is found between household energy consumption and user behavior \cite{Alsalemi2019is,ALMARRI2018,SARKODIE2019,ZHOU2016,ISHAK2016}. Every day, new appliances are installed and used in households resulting in an incredible rise of power demand. Monitoring the power usage of these appliances is dependably the initial step towards energy preserving. From this point of view, understanding and controlling user consumption behavior is also a key parameter to help householders reduce energy costs. According to the ongoing evolution of the Internet of Things (IoT), the use of smart meters for monitoring electricity consumption is expanding exponentially \cite{CSOKNYAI2019,VANAUBEL2019,AVANCINI2019}. The up-and-coming generation of energy saving systems should be more effective, easy to follow and more challenging in order to improve end-user behavior.

As of now, as well as other research areas, energy, environment and sustainable development research topics are encountering the urgency of developing openly accessible datasets. In fact, power consumption datasets progressively come to be more consistent when estimating the precision of power monitoring techniques and perceiving how good they may behave under realistic circumstances. Consequently, checking the precision of outputs in real scenarios is critical in this research area \cite{Latif2017,Moletsane2018,Mehar2018}. Moreover, simulated database does not reasonably fit realistic datasets as \enquote{an experimental database or repository would ordinarily have unpredictable and unexplained complication nature that is laboriously anticipated and most of the time can be laboriously hard to manage \cite{RAE2017,RAMIREZ2017}}. On that account, Energy scientists have proved that it is important to have open access databases that provide aggregated power consumption as well as appliance based consumption for the various devices that constitute the overall consumption \cite{Alsalemi8959214}.

Moreover, end-users behavior is responsible of wasting more than 20\% of the total energy consumed in buildings, and hence it is a key element in energy consumption \cite{White2019,URGEVORSATZ201585,ALMARRI20173464}. Therefore, it is of paramount importance to: (i) design real consumption datasets; (ii) deploy novel platforms and smart-meters to collect granular and appliance-specific data, which have the means or incentive for sharing consumption data with end-users; and (iii) develop tools that help end-users in understanding their energy consumption footprints, such as innovative visualizations, and further implement novel strategies that help them in improving their behavior and reduce wasted energy \cite{Paone11040953}, e.g. via deploying recommender systems.
In this context, seeking to study how to save energy and understand power consumption behaviors in buildings, various datasets have been collected globally. 
They provide a big amount of information and create an immense quantity of readings about daily power usage and user behavior. Therefore, the use of machine learning (ML) algorithms becomes essential for handling large-scale datasets and extracting meaningful features from collected data. This can assist in many applications including forecasting power demand, energy efficiency and electricity preserving, appliance recognition, cost prediction, and other things in relation to energy usage \cite{LIU2019,GUO2018,NGO2019}. In this respect, any ML algorithm for building power consumption deals with information drawn from smart meters and solar panels during the different periods of the day. This huge quantity of data including multivariate time series is of utmost importance to ML algorithms because future usage can be effectively anticipated \cite{XU2019,CHOU2018,WANG2019}. In other words, ML algorithms can forecast such data and help in developing energy efficiency frameworks proficiently.

Extremely inspired by the rising relevance of public datasets, we opted to give a general review of up to 31 existing datasets in the field of building power consumption. Presently, householders, firms and public authorities are confronting difficulties to guarantee the energy efficiency and reduce usage costs. The use of a large number of appliances increases the energy demand, cost and carbon emissions as well. In this context, we present in this paper a deep overview of various power consumption datasets, their taxonomy and classification. The taxonomy is adopted to examine existing datasets, what kind of information they provide, their applications, their benefits and their limitations. To the best of our knowledge, this is the first framework that provides a comprehensive and universal survey of building energy consumption datasets, their applications and future trends. Following, discussions and important findings are presented via analyzing and comparing the features of existing datasets, their data collection platforms and related modules. Moreover, a novel dataset, namely Qatar university dataset (QUD) is proposed to answer the challenges and issues raised from the analytical study. Afterward, valuable future orientations to improve the quality of power consumption datasets and enrich their content are discussed, in which a set of recommendations to use novel hardware devices and platforms are described. Moreover, future direction to improve datasets exploitation and hence improve energy saving are also also identified. To summarize, this paper presents a set of novel contributions, which can be listed as follows:

\begin{itemize}
\item Reviewing up to 31 building power consumption datasets, describing their properties and highlighting their pros and cons via adopting a multi-perspective comparison based on various parameters.
\item Proposing a taxonomy of building power consumption datasets to assess the existing repositories based on their applications an characteristics.
\item Analyzing data collection platforms used to record power consumption datasets and related modules used for data transmission, data storage and privacy concerns. 
\item Presenting a novel dataset called QUD that responds to various issues raised in the analysis of state-of-the-art datasets. QUD can be used for different applications, among them detecting of anomalous power consumption.
\item Providing a list of valuable future orientations for (i) improving datasets collection mainly through the use of novel hardware platforms, and (ii) improving datasets exploitation via adopting innovative tools such as visualization strategies and explainable recommender systems. 
\end{itemize}

The rest of this paper has been organized into four sections. Section \ref{sec2} reviews up to 31 existing building power consumption datasets and describes their usage contexts, properties, advantages and limitations. Section \ref{sec3} presents a comprehensive discussion about the different characteristics of existing power consumption datasets. In addition, a novel dataset called QUD is presented which presents new functionalities. In Section \ref{sec4}, challenging orientations and future directions that should be followed in order to improve datasets collection and enhance datasets exploitation are described. Section \ref{sec5} concludes the paper with a set of proposals for improving the quality of power consumption datasets and highlights future works. Finally, a list of abbreviations and nomenclatures used in this paper is presented in the Appendix.

\section{Overview of building power consumption datasets } \label{sec2}
\begin{figure}[t!]
\begin{center}
\includegraphics[width=16.9cm,height=6.2cm]{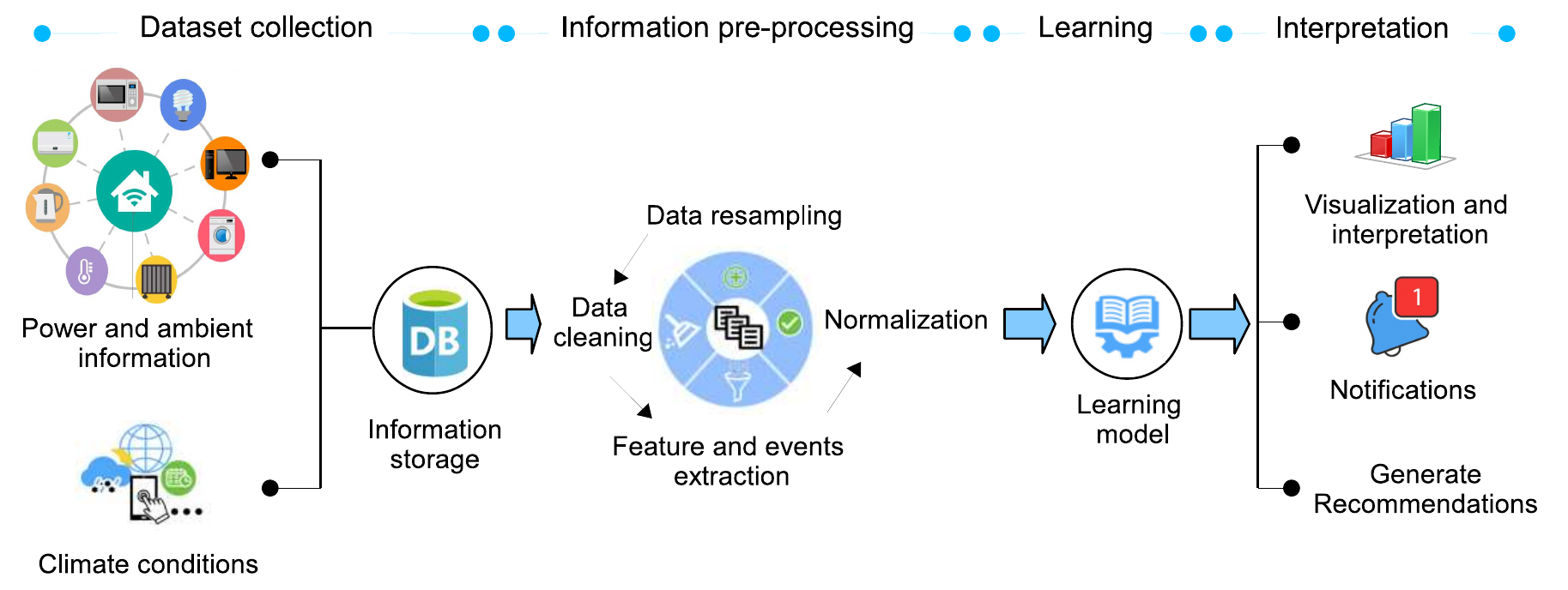}
\end{center}
\caption{General framework of an energy efficiency system.}
\label{EnergyFramework}
\end{figure}

Several datasets can be found in literature and each one has its specific characteristics, making it difficult to select a database for treating energy efficiency issues. To this end, this work makes a deep comparison between all datasets based on various specifications, such as the period and region of collection, sampling rate, number of monitored houses, number of deployed sub-meters, collected features and release date. As a matter of fact, existing realistic datasets are divided into two major groups; appliance-level datasets versus aggregated-level based databases. The first group class provides sub-meter readings of appliance-by-appliance consumption. This kind of data is used for various applications, including energy saving \cite{ALSALEMI2019ae}, appliance recognition \cite{Ruzzelli2010}, occupancy detection \cite{Gao2018,Sala2016} and preference behavior \cite{AHMADI2017,Franco2018}. The second class group focuses on collecting overall consumption profiles of different buildings. It can be employed for energy disaggregation, energy efficiency, and further predicting energy consumption. Figure \ref{EnergyFramework} illustrates a flowchart of a dataset collection process along with its associated modules, required to pre-process, analyze and interpret power consumption patterns. This is a general representation that can be used for different applications. 


\subsection{State of the art of existing datasets} \label{2.1}
To fit realistic scenarios of daily power usage and test energy efficiency solutions, scientists and specialists of smart energy monitoring systems need power consumption databases, in which developed algorithms can be evaluated in advance. Different databases have been collected and shared publicly. Under this section we review up to 31 power consumption datasets that are proposed in literature in addition to our novel dataset named QUD.  We specify briefly the characteristics of each dataset and registered features in terms of current (I), voltage (V), active power (P), reactive power (Q), apparent power (S), normalized power (Np), energy (E), frequency (f), phase angle ($\phi$), power factor (pf), energy cost (EC), weather (Wt), Temperature (T), humidity (H), occupancy (O) and light level (L).

In \cite{HES2012,IHEPCDS2013,UMAss2012,Pereira2014,REFIT2015,Dataport2015}, large-scale datasets are formed, namely, HES, IHEPCDS, UMSM, SustData, REFIT and Dataport, respectively. While HES, IHEPCDS, UMSM and Dataport assembled energy consumptions patterns at a minutely level, SustSata and REFIT reported power usage profiles over intervals in seconds. All these databases provide consumption records at the appliance-level for long periods of monitoring. For example; in HES and UMSM data are raised for a period of one year, in REFIT and SustData energy patterns are accumulated for 213 days and 1114 days, respectively. Further, different features are gathered during the experimental campaign, such as I, V, P, Q, S f and T. REFIT has also the particularity of providing EC in \$. Dataport repository \cite{Dataport2015} is also quite similar to UMSM database, since it captures energy usage at the same sampling intervals of 239 households but for a short collection period of two months. Dataport repository is also quite similar to UMSM database, since it captures energy usage at the same sampling intervals of more than 1200 households for a long  collection period, which is more than 4 years.

In \cite{OCTES2013}, OCTES is proposed, which is similar to REFIT. It records P,  $\phi$ and EC (\$). In addition, data are collected in a shorter investigation period. A bigger examination size and information are recorded at a comparable rate to REFIT. It lists the power consumption of each house;
nonetheless, other pieces of information about the houses are not provided except their geological position. The case study specfied in this work depicts the utilization of a sauna in one home; as though, this data isn't shared publicly. Consequently, a presumption should be put with regards to the energy usage. In addition to power consumption, REFIT provides also readings about temperature, light, and motion patterns expanded with dwelling reviews specifying;  size, age, warming sort, isolation,fabrication type and details about the tenants or occupants, job description and age.

In \cite{TraceBase2012}, Tracebase database includes power consumption patterns of various devices, which enables to examine disaggregation. The readings are collected at a sampling rate of 1 second. This dataset can be utilized for energy efficiency applications. However, it can not be employed for appliance recognition, preference detection or energy disaggregation since no data are provided about the devices being investigated and their properties. It gathers data of 43 distinct appliances, in which every one has various recordings from several days and several households. Furthermore, date and time records, P and Np  are provided at a sampling frequency of 8 seconds.

In \cite{AMPds1-2013,AMPds2-2016,ECB2014,PSD2012}, AMPds1, AMPds2, ECB and PSD are proposed, respectively,  which are minutely power datasets. Overall, AMPds1 and AMPds2 repositories are deemed as largely used databases, which compiled information of one and two years, accordingly, with a sampling rate of 1 min. In fact, energy consumption of 11 appliances is observed using 21 sub-meters. On the other side, ECB that provides electricity consumption benchmarks of 25 domestic residents located in Victoria State in south-eastern Australia is released. Consumption patterns were extracted from the aggregated circuit and for individual appliances over a duration of two years and at a sampling rate of 30 min. Further, consumption footprints of device-event labels from 10 homes in Austin, USA, were assembled.

In \cite{MEULPv1} and \cite{MEULPv2} authors released MEULPv.1 and MEULPv.2 datasets, respectively. MEULPv.1 gives energy consumption readings of 12 Canadian households. Data were recorded at 1-min sampling rates at both the aggregated and appliance levels. A total of 8 appliances are monitored during the data collection process. Meanwhile, MEULPv.2 provides one year monitoring of 23 households using a sampling rate of 1 min that designates aggregated and appliance-based consumptions as well.

In \cite{RAE2017,GREEND2014}, RAE and GREEND databases are proposed, in which data are collected at a frequency of 1 Hz. The RAE is the initial version of an energy consumption repository that includes 1 Hz recordings for aggregated and sub-metered levels of two households. Besides power information, T and H records from a house's indoor regulator are incorporated. On the other side, GREEND is proposed to describe detailed energy consumption patterns collected through an experimental campaign via assessing electricity usage of various individual appliances in Austria and Italy. During the collection campaign, eight households are monitored, where each one contains up to nine different individual devices. The power usage patterns at a device-level are gleaned at a resolution of 1 Hz through a period of six months.

In \cite{ECOD2014,IAWE2013,DRED2015}, ECO, IWAE and DRED that capture energy information at 1 Hz sampling intervals are nominated, accordingly. ECO is an entire measurement campaign managed in order to collect comprehensive information of consumption patterns in six Swiss homes through an eight months duration. During the collection campaign, I, V, and $\pi$ are collected from aggregated circuits and a set selected appliances at a sampling frequency of 1 Hz. Through the IAWE campaign, measurements were performed in a pilot household with three floors in Delhi in order to measure power, water and environmental profiles. Data are collected for a duration of 73 days from May to August 2013. In addition, 33 sub-meters are deployed through the whole house. DRED is publicly launched to capture energy, occupancy patterns and environmental data of one pilot house in the Netherlands. Sensor units are installed to measure aggregated energy consumption and appliance level electricity usage. In fact, 12 different domestic appliances are sub-metered at sampling intervals of 1 min while 1 Hz sampling rates are used to gather aggregated consumption.

In \cite{DISEC2018}, DISEC is launched, in which various data are collected for 19 apartments at an Indian faculty housing complex during 284 days. Different features, such as P and Wt, are collected in a 30 seconds sampling intervals and then aggregated to 15 min, 30 min and 60 min intervals. As well, Wt variations are updated through measuring atmospheric conditions from nearly station measurements.

In \cite{CRHLL2013} and \cite{HUE2019}, two hourly electricity consumption datasets are proposed. The first one called CRHLP includes  energy patterns of 16 residential and commercial buildings monitored at every hour for a period of one year. Additionally, solar radiation and meteorological records are also collected. The second one, namely HUE, captures long-term energy usage profiles from five households with a sampling frequency of 1 hour. Furthermore, while device-level consumptions from house 1 are collected for a period of two years with sampling intervals of one minute, data from house 2 are extracted for a one year period with a resolution rate of 1 Hz. In \cite{UK-DALE2015}, UK-DALE is proposed, which summarizes the current and voltage profiles of three houses at sampling intervals of 16 KHz and two houses at sampling frequencies of 1 Hz. Moreover, patterns of individual devices of five other households are collected at a sampling rate of 6 seconds for various periods varying from 39 to 655 days.

In \cite{REDD2011,BLUED2012,BLOND2018}, REDD, BLUED and BLOND datasets are proposed. Energy consumption records are captured at a sampling frequency of more than 10 kHz. The monitoring process, by contrast, is conducted for only a few weeks. For example, in REDD, six households are monitored, where the aggregated electricity consumption is measured at a sampling rate (15 KHz). Also, electricity consumption reviews of up to 24 devices are monitored at sampling intervals of 0.5 Hz. Furthermore, load patterns of other 20 appliances are observed at a frequency of 1 Hz while BLUED resumes the current and voltage readings of an individual household in Pittsburgh, Pennsylvania, USA. Data are listed at a sampling frequency of 12 kHz over a period of one week. For BLOND, it aims to capture continuous power consumption data. It delivers voltage and current records at the aggregated and device levels. This database includes data from 53 devices that represent 16 appliance groups. It englobes two main repositories; (i) BLOND-50 that in turn has consumption data obtained at sampling intervals of 50 kSps for grouped circuits and 64 kSps for individual devices; and (ii) BLOND-250 that entails usage patterns for a period of 50 days gathered using sampling rates of 250 kSps at the aggregated-level and 50 kSps at the appliance-level.

In \cite{PLAID2014}, PLAID expresses power consumption profiles for more than 56 specific domestic equipments that represent about 11 appliance categories. Data are captured at a sampling frequency of 30 kHz that is judged among the highest resolution frequency used in existing building power consumption datasets when collecting load profiles. In addition, energy consumption information is captured for a period of three months during the summer of 2013 and the measurement campaign has been carried on in Pittsburgh, Pennsylvania, USA.

ACS-F1 \cite{ACS-F1-2013} and BERDS \cite{BERDS2013} datasets that monitor load patterns at a comparable sampling rate are proposed. ACS-F1 records the amount of energy used in a set of households at an appliance-level. In this context, electricity sub-meters were employed to measure the energy consumption of 100 house devices that represent 10 appliance classes. Power sub-metering is managed at sampling intervals of 10 seconds for a period of only one hour. This database is especially suitable for appliance recognition applications. On the other side, BERDS collects energy consumption outlines at 20 seconds sampling rates for a period of one year. 

In addition, QUD is presented in this framework, which is based on an appliance-based collection campaign. It can be used for different for different purposes, such as the energy saving, anomaly detection and energy demand prediction. QUD is collected using a system that incorporates sub-metering modules registering power consumption footprints in terms of P and other indoor climate conditions, including O, T, H and L. The data are recorded with sampling intervals ranging from 3 seconds to 30 min. The collection process will be spread over a one year period, while three months of data recording have been already completed. 

\begin{figure}[t!]
\begin{center}
\includegraphics[width=16.3cm,height=5.6cm]{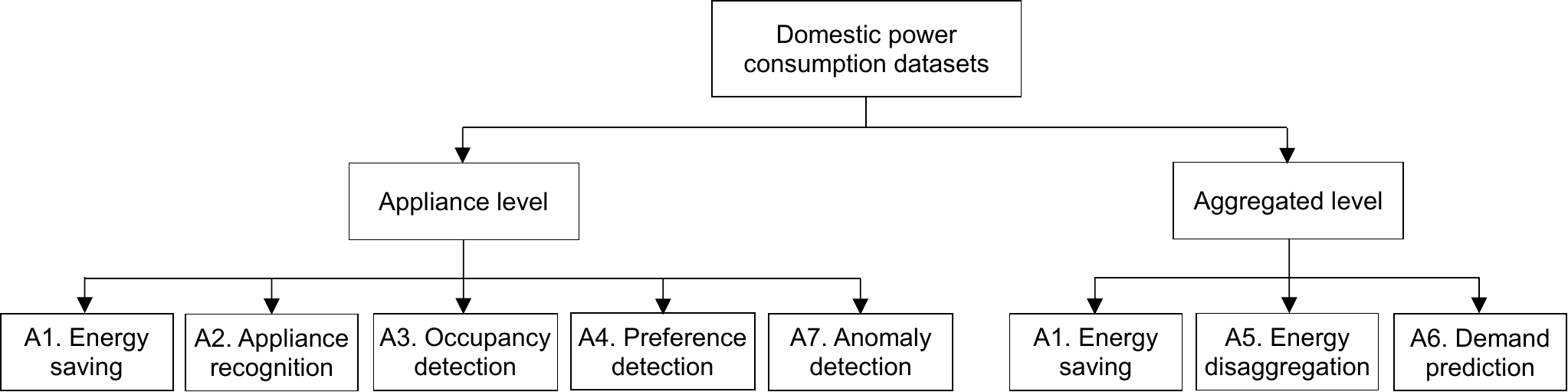}
\end{center}
\caption{General taxonomy of existing building power consumption datasets.}
\label{taxonomy}
\end{figure}

\subsection{Taxonomy}
Power consumption datasets are split into two main groups: Appliance-level versus aggregated-level. The first one traces power consumption arrangements of individual devices. The second one provides the whole power consumption of households. Datasets can also be classified based on different aspects including application purposes or the nature of buildings, where data are acquired among which households, commercial buildings, academic buildings, industrial, etc. Figure \ref{taxonomy} details the global taxonomy of various building power consumption datasets found in the literature.

\subsection{Applications (A)}
Using detailed power consumption readings and based on the nature of data collection procedures at appliance or aggregated levels, existing datasets could be exploited for various applications including, but not restricted to, energy saving, appliance recognition, occupancy detection, user preference detection, abnormal detection, energy disaggregation and energy demand prediction.

\vskip2mm

\noindent \textbf{A1. Energy saving:} 
Investigating the building sector in terms of energy saving which is a principal element of its environmental and financial effects is of utmost importance. Consequently, energy saving is the most popular application of building power consumption datasets  \cite{Alsalemi2019IntelliSys,Sardianos2019,Himeur2020icict}. It can effectively reduce energy bills and decrease carbon dioxide emissions. It is made out of the following four stages: (1) the dataset collection stage, in which information is reaped from various sources, including energy sub-meters, ambient condition sensors and climate sources. The information gathered from these heterogeneous sources is saved in a specific dataset; (2) the pre-processing step, in which the information stored in the first step is pre-processed before utilizing various ML strategies. the pre-processing includes data cleaning, data resampling, features and events extraction and normalization; (3) The learning stage, in which ML algorithms are utilized to learn functions and models; and (4) The adoption of visualizations and recommendations phase, in which visualization tools are first adopted to provide end-users with interpretation of their consumption patterns. Following, specific recommendations or directives are derived in order to promote energy efficiency behaviors. Since the energy saving application is very relevant, we focus in this paper on studying how to improve systems developed in this direction along with related applications.

\vskip2mm

\noindent \textbf{A2. Appliance recognition:}
Appliance recognition systems can help detecting operating conditions of devices using collected power usage patterns, and thoroughly recognizing the nature of each appliance \cite{ROSSIER2017}. In \cite{Lee2010ARU}, a model was designed to detect the device activity and then to associate activities with devices using collected data. Analyzing power signals and checking relations among activities can assist detecting unattended devices, which use energy power without taking part the domestic's activities. In \cite{Kahl:2017}, in order to fit realistic conditions, experiments are usually conducted on a set of building power consumption databases, such as ACS-F1, PLAID, BLUED and UK-DALE.
\vskip2mm

\noindent \textbf{A3. Occupancy detection:}
Solutions presented in this area detect individuals' occupancy in each specific part of a building based on power consumption profiles, as well as other environmental specifications, such as the temperature, humidity, luminosity and carbon dioxide emissions \cite{Sardianos2020iciot}. Dataset patterns are inspected before using ML approaches to derive the occupancy of the monitored part. Generally, occupancy is detected in two stages; (i) the presence or absence of individuals is investigated; and (ii) the number of individuals in the monitored building/room is then calculated \cite{WEI2019,AHMAD2018,CHEN2018260}. In \cite{Vafeiadis2017}, a set of ML models as well as their boosting forms are developed and tested to detect occupancy using collected data from the AMPds2 measurement campaign.  
 
\vskip2mm

\noindent \textbf{A4. Preference detection:} 
Methods described in this class deal with evaluating individual preferences through analyzing energy usage profiles. Most approaches treat the thermal comfort, although there are other arrangements that address visual comfort. Works released in this area investigate information-driven methodologies from an ML point of view and yielded arrangements that determine the preferences (e.g. the habits related to appliance usage) even through getting reports from individuals, i.e. information labeling or via observing the historic behavior of end-users to construe (in a straightforward manner) their consumption priorities or contexts that satisfy their well-being \cite{AHMADI2017,KHASHE2015}. 

\vskip2mm

\noindent \textbf{A5. Energy disaggregation:} 
Energy disaggregation is the issue of segregating the overall power consumption record into particular signals, in which each one represents an individual consumption of each electrical device \cite{TANG201742,BRESCHI2018,AIAD2018,MIYASAWA2019}. This is valuable since getting separated power consumption of each appliance helps individuals to save energy and provide consumers with indexes on how to make appropriate actions \cite{Himeur2020iscas}. Most of existing energy disaggregation frameworks resolving the problem of non-intrusive load monitoring (NILM) attempt to segregate the overall energy consumption without utilizing separate meters for each appliance \cite{LIU2018209,WANG2018134,LIU20181106,HENRIET2018268,HOSSEINI20171266}. For this specific application, REDD, BERDS, REFIT, AMPds1 and AMPds2 datasets are reputed among the famous repositories used for energy disaggregation. 

\vskip2mm

\noindent \textbf{A6. Demand prediction:} 
ML algorithms generate precise power demand forecasts and they can be selected by public authorities and project managers instrumenting energy-efficiency procedures \cite{WANG2019e,BOURDEAU2019,AHMAD201,HONG2019,HERREA2019}. For domestic households, academic and industrial buildings, if the power demand could be predicted using ML strategies, directives and mechanisms that should be followed in advance can be established with a view of reducing load consumption of equipments and appliances inside these infrastructures\cite{LI2018404,VILLCAPOZO2019,LOPEZOCHOA2019,THONIPARA2019}. Moreover, even if most the above presented databases (Section \ref{2.1}) are used for energy forecasting, we can find in the literature other datasets that are only designed for the specific problem of load and energy price forecasting, such as GEFCom2012 \cite{HONG2014357} and GEFCom2014 \cite{HONG2016896}. 

\vskip2mm

\noindent \textbf{A7. Anomaly detection:} 
With the progressive widespread use of smart-meters and smart sensors to monitor load usage in households, the utilization of power consumption observations as a solution to detect abnormal usage of energy is absolutely fascinating. Specifically, early detection approaches can be deployed to identify a large set of failures. In addition, recent works illustrate that for example, anomalous in lighting appliances can be responsible of 2--11 \% of the whole power consumption of households and and commercial structures \cite{Cui2017}. Furthermore, detecting faults or anomalies can permit analysts to comprehend energy consumption behavior of end-users and to be conscious of unpredictable energy usage values \cite{SEEM2007,KHAN2013}. Various data mining approaches have been explored and deployed to detect  anomalous events during energy usage process \cite{JANETZKO2014,MA2017,ARAYA2017,Nordahl2017,Qiu2018,Weng2019}. In addition, it is worthy to mention that there is an absence of annotated datasets dedicated to power consumption anomaly detection.

However, in order that a dataset could be correctly and efficiently used for a specific application, it should respect some specific requirements. For energy disaggregation, datasets should include both aggregated and appliance-level consumption fingerprints to compare the results obtained from disaggregation solutions with individual patterns. To conduct a user preference detection or even an occupancy detection, datasets should encompass appliance-level power consumption because it is difficult even impossible to infer user preferences from aggregated data. In addition, for occupancy detection, it is also required that consumption and ambient condition should be gleaned from individual appliances and from various parts of the building. For anomaly detection, it is of utmost importance that it includes labels annotating normal and anomaly consumption footprints to train developed algorithms. Lastly, for energy demand prediction, collecting power consumption at appliance-level or aggregated level will be appreciated, however, the collection period should be long to be useful.

\subsection{Characteristics comparison of existing datasets} 
Aiming to extract representative outputs and relevant interpretations, a deep comparison study of existing building power consumption datasets is conducted in this section. Various dataset properties are investigated, which have a great importance when collecting data for developing energy efficiency solutions. Table \ref{datasets_comp} presents a comparative investigation of existing power consumption datasets. The analysis is built based on various characteristics that were collected in each dataset, including the region and period of collection, number of monitoring houses, number of monitoring appliances per house, collected features, sampling rate and release year. Additionally, we check and compare collected features for each database.

\begin{table}[t!]
\caption{Features comparison of existing building power consumption datasets.}
\label{datasets_comp}
\begin{center}

\begin{tabular}{l|l|l|l|l|l|l|l|l|l}
\hline
{\tiny \#} & {\tiny Acronym} & {\tiny Country} & {\tiny Period} & {\tiny %
\#Homes} & {\tiny \#sub-meters} & {\tiny Features} & {\tiny Sampling rate} & 
{\tiny Applications} & {\tiny Release} \\ 
&  &  &  &  &  &  &  &  &  \\ \hline
{\tiny 1} & {\tiny REDD \cite{REDD2011}} & {\tiny Massachusetts, USA} & 
{\tiny 119 days} & {\tiny 6} & {\tiny 24} & {\tiny I, V, P} & {\tiny 3 sec}
& {\tiny A1,A5} & {\tiny 2011} \\ \hline
{\tiny 2} & {\tiny HES \cite{HES2012}} & {\tiny England, UK} & {\tiny 1 year}
& {\tiny 26/251} & {\tiny 23} & {\tiny I, V, P, T} & {\tiny 10 min} & {\tiny %
A1,A6} & {\tiny 2011} \\ \hline
{\tiny 3} & {\tiny IHEPCDS \cite{IHEPCDS2013}} & {\tiny Paris, France} & 
{\tiny 47 months} & {\tiny 1} & {\tiny 3} & {\tiny I, V, P, Q} & {\tiny 1 min%
} & {\tiny A1,A6} & {\tiny 2012} \\ \hline
{\tiny 4} & {\tiny UMSM \cite{UMAss2012}} & {\tiny Massachusetts, USA} & 
{\tiny 1 year} & {\tiny 400} & {\tiny 8} & {\tiny I, V, P, f, S} & {\tiny 1
min} & {\tiny A1} & {\tiny 2012} \\ \hline
{\tiny 5} & {\tiny Dataport \cite{Dataport2015}} & {\tiny USA} & {\tiny 4
years} & {\tiny +1200} & {\tiny 70} & {\tiny P} & {\tiny 1 min} & {\tiny A1}
& {\tiny /} \\ \hline
{\tiny 6} & {\tiny MEULPv1 \cite{MEULPv1}} & {\tiny Canada} & {\tiny 1 year}
& {\tiny 11} & {\tiny 8} & {\tiny P} & {\tiny 1 min} & {\tiny A1,A5} & 
{\tiny 2012} \\ \hline
{\tiny 7} & {\tiny BLUED \cite{BLUED2012}} & {\tiny Pennsylvania, USA} & 
{\tiny 1 week (Oct)} & {\tiny 1} & {\tiny Agg} & {\tiny I, V, switch events}
& {\tiny 12 KHz} & {\tiny A2,A5} & {\tiny 2012} \\ \hline
{\tiny 8} & {\tiny TraceBase \cite{TraceBase2012}} & {\tiny Darmstadt,
Germany} & {\tiny N/A} & {\tiny 15} & {\tiny 158 (43 classes)} & {\tiny P, Np}
 & {\tiny 1-8 sec} & {\tiny A1,A} & {\tiny 2012} \\ \hline
{\tiny 9} & {\tiny PSD \cite{PSD2012}} & {\tiny Austin, USA} & {\tiny 1 week}
& {\tiny 10} & {\tiny n/a } & {\tiny P} & {\tiny 1 min} & {\tiny A2} & {\tiny %
2012} \\ \hline
{\tiny 10} & {\tiny CRHLL \cite{CRHLL2013}} & {\tiny USA} & {\tiny 1 year} & 
{\tiny 16} & {\tiny 10} & {\tiny P} & {\tiny 1 hour} & {\tiny A1,A6} & 
{\tiny 2013} \\ \hline
{\tiny 11} & {\tiny IAWE \cite{IAWE2013}} & {\tiny New delhi, India} & 
{\tiny 73 days (May--Aug)} & {\tiny 1} & {\tiny 33 (10 classes)} & {\tiny I,
V, P, f, S, E, $\Phi $} & {\tiny 1 Hz} & {\tiny A1,A6} & {\tiny 2013} \\ 
\hline
{\tiny 12} & {\tiny ACS-F1 \cite{ACS-F1-2013}} & {\tiny Switzerland} & 
{\tiny 1 hour (2 sessions)} & {\tiny /} & {\tiny 100 (10 types)} & {\tiny I,
V, P, Q, f, $\Phi $} & {\tiny 10 sec} & {\tiny A2} & {\tiny 2013} \\ \hline
{\tiny 13} & {\tiny AMPds1 \cite{AMPds1-2013}} & {\tiny Vancouver, Canada} & 
{\tiny 1 year} & {\tiny 1} & {\tiny 21} & {\tiny I, V, P, Q, S, pf, F } & 
{\tiny 1 min} & {\tiny A1,A2,A5} & {\tiny 2013} \\ \hline
{\tiny 14} & {\tiny BERDS \cite{BERDS2013}} & {\tiny Berkely, USA} & {\tiny %
1 year} & {\tiny /} & {\tiny 4 groups} & {\tiny P, Q, S} & {\tiny 20 sec} & 
{\tiny A1} & {\tiny 2013} \\ \hline
{\tiny 15} & {\tiny ECODS \cite{ECOD2014}} & {\tiny Switzerland} & {\tiny 8
months} & {\tiny 6} & {\tiny /} & {\tiny I, V, $\Phi $} & {\tiny 1 Hz} & 
{\tiny A1} & {\tiny 2014} \\ \hline
{\tiny 16} & {\tiny ECB \cite{ECB2014}} & {\tiny Australia} & {\tiny 2 years}
& {\tiny 25} & {\tiny Aggregated} & {\tiny P} & {\tiny 1 Hz} & {\tiny A5,A6}
& {\tiny 2014} \\ \hline
{\tiny 17} & {\tiny PLAID \cite{PLAID2014}} & {\tiny USA} & {\tiny 3 months
(Summer)} & {\tiny 11} & {\tiny 60} & {\tiny I, V} & {\tiny 30 KHz} & {\tiny %
A2} & {\tiny 2014} \\ \hline
{\tiny 18} & {\tiny SustData \cite{Pereira2014}} & {\tiny Portugal} & {\tiny %
1144} & {\tiny 50} & {\tiny 24} & {\tiny I, V, P, Q, S} & {\tiny 2 sec / 10
sec} & {\tiny A1} & {\tiny 2014} \\ \hline
{\tiny 19} & {\tiny AMPds2 \cite{AMPds2-2016}} & {\tiny Vancouver, Canada} & 
{\tiny 730 days} & {\tiny 1} & {\tiny 21} & {\tiny I, V, P, S, F, pf} & 
{\tiny 1 min} & {\tiny A1,A2,A5} & {\tiny 2014} \\ \hline
{\tiny 20} & {\tiny UK-DALE \cite{UK-DALE2015}} & {\tiny England, UK} & 
{\tiny 655 days} & {\tiny 4} & {\tiny 5 (H4), 53 (H1)} & {\tiny P,
Aggregated P} & {\tiny 6 sec/ 6 KHz (Agg) } & {\tiny A1} & {\tiny \ 2015}
\\ \hline
{\tiny 21} & {\tiny DRED \cite{DRED2015}} & {\tiny Netherland} & {\tiny 6
months (Jul--Dec) } & {\tiny 1} & {\tiny 13} & {\tiny P, T, H, Ws, Pr, Agg}
& {\tiny 1 min / 1 Hz (for Agg)} & {\tiny A1,A3,A4} & {\tiny 2015} \\ \hline
{\tiny 22} & {\tiny GREEND \cite{GREEND2014}} & {\tiny Italy \& Austria} & 
{\tiny 6 months} & {\tiny \ 8} & {\tiny 9} & {\tiny P} & {\tiny 1sec Hz} & 
{\tiny A1,A5} & {\tiny 2015} \\ \hline
{\tiny 23} & {\tiny REFIT \cite{REFIT2015}} & {\tiny England, UK} & {\tiny %
213 days} & {\tiny 20} & {\tiny 9, Agg} & {\tiny P, pf, T, O, L, EC ($\$$}%
{\tiny )} & {\tiny 8 sec} & {\tiny A5} & {\tiny 2015} \\ \hline
{\tiny 24} & {\tiny OCTES \cite{OCTES2013} } & {\tiny Scotland, UK} & {\tiny %
4 - 13 months} & {\tiny 33} & {\tiny Agg} & {\tiny P, EC $(\$)$} & {\tiny 7
sec} & {\tiny A5} & {\tiny 2015} \\ \hline
{\tiny 25} & {\tiny COOLL \cite{COOLL2016}} & {\tiny France} & {\tiny 2 hours%
} & {\tiny 1} & {\tiny 46 (12 groups)} & {\tiny I, V} & {\tiny 100 KHz} & 
{\tiny A2} & {\tiny 2016} \\ \hline
{\tiny 26} & {\tiny MEULPv2 \cite{MEULPv2}} & {\tiny Canada} & {\tiny 1 year}
& {\tiny 23} & {\tiny 5 groups} & {\tiny P} & {\tiny 1 min} & {\tiny A1} & 
{\tiny 2017} \\ \hline
{\tiny 27} & {\tiny RAE \cite{RAE2017}} & {\tiny Canada} & {\tiny 72 days} & 
{\tiny 1} & {\tiny 24} & {\tiny O, V, P, Q, S, f, E} & {\tiny 1 Hz} & {\tiny %
A1,A6} & {\tiny 2018} \\ \hline
{\tiny 28} & {\tiny DISEC \cite{DISEC2018}} & {\tiny New Delhi, India} & 
{\tiny 284} & {\tiny 19} & {\tiny /} & {\tiny P, Wt} & {\tiny 30 sec/15, 30,
60 min (Agg)} & {\tiny A1,A5} & {\tiny 2018} \\ \hline
{\tiny 29} & {\tiny BLOND \cite{BLOND2018}} & {\tiny Germany} & {\tiny 213
days} & {\tiny 1} & {\tiny 53 (16 groups)} & {\tiny I, V, P} & {\tiny 6.4
kSps/54 kSps (Agg)} & {\tiny A1,A2} & {\tiny 2018} \\ \hline
{\tiny 30} & {\tiny HUE \cite{HUE2019}} & {\tiny B. Columbia, Canada} & 
{\tiny 3 years} & {\tiny 5} & {\tiny /} & {\tiny p} & {\tiny 1 hour, H1(1
min), H2 (1 Hz)} & {\tiny A1} & {\tiny 2019} \\ \hline

{\tiny 31} & {\tiny ENERTALK \cite{ENERTALK2019}} & {\tiny Seoul, South Korea} & 
{\tiny 122 days} & {\tiny 22} & {\tiny 1-7 (Agg)} & {\tiny p} & {\tiny 1 Hz,
15 Hz} & {\tiny A1,A5} & {\tiny 2019} \\ \hline
{\tiny 32} & {\tiny QUD } & {\tiny Doha, Qatar} & 
{\tiny 3 months--1 year} & {\tiny 3} & {\tiny 4} & {\tiny P, H, T, O} & 
{\tiny 3 sec--30 min} & {\tiny A1,A3,A4,A7} & {\tiny 2019} \\ \hline
\end{tabular}

\end{center}
\end{table}

\subsection{Data collection Platforms}
Data collection platforms used to glean big energy consumption fingerprints are significantly impacting the energy efficiency systems. Specifically, sensing devices and attached platforms have a big role in gathering and safely storing data in appropriate databases. In this line, in this subsection, we focus on inspecting different architecture platforms used in the literature to collect energy consumption datasets and their properties, including wireless capability, data logging process and and data storage. In addition, because of the nature of collected data and their public access capability, privacy concerns are of utmost importance when producing datasets. Specifically, transmitting and sharing individuals' real-time power usage footprints and further their identities are probably quite harmful. To that end, it is important to investigate if the connections to the servers are secure or not in the presented dataset platforms. It is worthy to mention that in this section we focus on analyzing hardware architectures and related modules for only the datasets from Table \ref{datasets_comp}, which present a description of their implemented platforms.

In \cite{Pereira2014}, a power consumption monitoring and feedback platform is deployed, which is based on the use of sensors and a notebook for recordings data, storing them on MongoDB database, performing calculations and providing feedback to end-users. 
In \cite{REFIT2015}, readings from several smart appliances are collected and transmitted using a commercial communication gateway called Vera3 smart
home controller. The latter uses an encryption protocol to transmit data before their storage in a MySQL database.
In \cite{AMPds2-2016}, electricity footprints are gleaned using industrial meters and transmitted using a commercial platform named Obvius AcquiSuite EMB A8810, which includes many feature, among them the security provision. After that, they are stored offsite on a MySQL database server.  
In \cite{GREEND2014}, platforms based on Raspberry Pi or BeagleBone along with a Plugwise Basic kit4 are used, in which collected data from sensing outlets are transmitted via a Zigbee network. Collected data are then stored on via a remote storage on a MySQL server without considering privacy concerns.
In \cite{REDD2011}, a wireless plug monitoring device with an off-the-shelf system are used to collect power consumption data before transmitting them to central server. To keep the privacy of end-users, REDD dataset has focused only on hiding the identity of end-users and without deploying any secure protocol for data transmission.

In \cite{RAE2017}, power consumption readings of several appliances are wirelessly gathered using a data acquisition platform based on a Raspberry Pi 2B. Then, data are locally stored on an USB drive.
In \cite{UK-DALE2015}, a Nanode platform is used to wirelessly collect consumption data from individual appliance monitors and current transformers. Following, gleaned data are stored in a Nanode base station. It is worthy to mention privacy issues have not been considered.
In \cite{ENERTALK2019}, consumption records are acquired using a commercial plug, namely ENERTALK PLUG, which includes a microcontroller unit to process and save them in a device
storage unit. After that, they are wirelessly transmitted to a data collector server. Finally, data are saved on a NoSQL Hadoop database server.

In this framework power consumption is measured using submeters components such as NodeMCU and SEN-11005 current transformer. Furthermore, occupancy patterns, luminosity, temperature and humidity data are also recorded using smart-sensors and then transmitted wirelessly using Raspberry Pi 4 Model B platform. The latter includes a No SQL CouchDB server that is used to store the gathered data using the JavaScript Object Notation (JSON). JSON represents a vastly used text format for data exchange, which keeps data structure without adding notation overhead. Table \ref{Platforms} summarizes the properties of hardware platforms used to collect different datasets, including wireless capability, data logging process, data storage and privacy consideration.

\begin{table}[t!]
\caption{Example of data collection platforms and their properties used in different datasets.}
\label{Platforms}
\begin{center}

\begin{tabular}{cccccc}
\hline
{\small Dataset} & {\small Platfom} & {\small Wireless} & {\small Data
logging} & {\small Data storage} & {\small Privacy} \\ 
&  & {\small capability} &  &  & {\small consideration} \\ \hline
\multicolumn{1}{l}{\small REDD} & \multicolumn{1}{l}{\small Laptop + smart
meters} & \multicolumn{1}{l}{\small yes} & \multicolumn{1}{l}{\small -} & 
\multicolumn{1}{l}{\small Hard drive} & \multicolumn{1}{l}{\small no} \\ 
\multicolumn{1}{l}{\small GREEND} & \multicolumn{1}{l}{\small %
Raspberry/BeagleBone} & \multicolumn{1}{l}{\small yes} & \multicolumn{1}{l}%
{\small JSON} & \multicolumn{1}{l}{\small MySQL server} & \multicolumn{1}{l}%
{\small no} \\ 
\multicolumn{1}{l}{\small SustData} & \multicolumn{1}{l}{\small Laptop +
smart meters} & \multicolumn{1}{l}{\small non} & \multicolumn{1}{l}{\small %
JSON} & \multicolumn{1}{l}{\small MongoDB server} & \multicolumn{1}{l}%
{\small no} \\ 
\multicolumn{1}{l}{\small REFIT} & \multicolumn{1}{l}{\small Vera3 smart
home controller Vera3} & \multicolumn{1}{l}{\small yes} & \multicolumn{1}{l}%
{\small JSON} & \multicolumn{1}{l}{\small MySQL server} & \multicolumn{1}{l}%
{\small yes} \\ 
\multicolumn{1}{l}{} & \multicolumn{1}{l}{\small + smart plugs} & 
\multicolumn{1}{l}{} & \multicolumn{1}{l}{} & \multicolumn{1}{l}{} & 
\multicolumn{1}{l}{} \\ 
\multicolumn{1}{l}{\small AMPDS} & \multicolumn{1}{l}{\small Obvius
AcquiSuite EMB A8810} & \multicolumn{1}{l}{\small non} & \multicolumn{1}{l}%
{\small SQL} & \multicolumn{1}{l}{\small MySQL server} & \multicolumn{1}{l}%
{\small yes} \\ 
\multicolumn{1}{l}{\small RAE} & \multicolumn{1}{l}{\small Raspberry Pi 2B +
sub-meters} & \multicolumn{1}{l}{\small yes} & \multicolumn{1}{l}{\small XML}
& \multicolumn{1}{l}{\small Local storage (USB drive)} & \multicolumn{1}{l}%
{\small no} \\ 
\multicolumn{1}{l}{\small UK-DALE} & \multicolumn{1}{l}{\small Nanode (Atmel
ATmega328P)} & \multicolumn{1}{l}{\small yes} & \multicolumn{1}{l}{\small %
JSON} & \multicolumn{1}{l}{\small Nanode base station} & \multicolumn{1}{l}%
{\small no} \\ 
\multicolumn{1}{l}{\small ENERTALK} & \multicolumn{1}{l}{\small ENERTALK PLUG%
} & \multicolumn{1}{l}{\small yes} & \multicolumn{1}{l}{\small Hadoop} & 
\multicolumn{1}{l}{\small NoSQL Hdoop database} & \multicolumn{1}{l}{\small %
no} \\ 
\multicolumn{1}{l}{\small QUD} & \multicolumn{1}{l}{\small NodeMCU +
Raspberry Pi 4B } & \multicolumn{1}{l}{\small yes} & \multicolumn{1}{l}%
{\small JSON} & \multicolumn{1}{l}{\small No-SQL CouchDB server} & 
\multicolumn{1}{l}{\small no} \\ 
& {\small + smart sensors} &  &  &  &  \\ \hline
\end{tabular}

\end{center}
\end{table}

\section{Discussion and important findings} \label{sec3}
\subsection{Discussion}
Under this framework, a large number of building power consumption databases have been described, reviewed and evaluated according to different parameters as indicated in Table \ref{datasets_comp}. In what follows, we derive pros and cons of each dataset, based on what has been discussed in the previous lines. This can adequately guides us to map recommendations for enriching and improving energy consumption databases. 

\begin{itemize}
\item The biggest databases in terms of length and period of study are UMSM, HES, SustData and REFIT. Otherwise, for the case of HES, the observing period is too short and the sampling frequency of 2 minutes is a bit big. The same for UMSM, where data are gathered at a sampling rate of 1 min. Therefore, these datasets are inadequate for energy disaggregation as it will be difficult to differentiate between individual devices and occurrences. In contrast, these two repositories provide properties data about monitored homes, among others, the nature of building, size and rooms number and occupants number. Moreover, even if SustData and REFIT use a sampling rate of 8 sec and 10 sec, this is still not enough when conducting a real-time monitoring.

\item In some databases, e.g. PLAID, REDD and BLUED, high frequency monitoring is proceeded for only a few number of houses. This draws upon the prerequisites of energy disaggregation, where comprehensive characteristics catching transitory behavior can be extracted when high frequency collection is explored.

\item The majority of databases were gathered in the USA and Canada, under a 120V voltage and European nations under 230V. It can be deduced from Table \ref{datasets_comp} that the existing databases are collected in 13 different countries which are located in four continents; inter alia, America, Europe, Asia and Australia. In this context, these real databases have been produced in distinct climate zones, which cover humid regions (UMSM, REDD, BLUED), humid semitropical (IAWE), marine west coast atmosphere (UK-DALE,HES, AMPds1, REFIT,  Tracebase, BLOND, IHEPCDS and OCTES), Mediterranean weather (BERDS) and arid zone (ECB ad QUD). 

However, no databases from Africa countries have been gathered under this investigation since there is no work in the literature who treat this topic in such countries. Moreover, to the best of the authors knowledge, QUD is the first dataset in the Middle East, where ordinarily 240 V voltage is used. Also, some collected particularities; for instance, the climate and environmental data depend on the location of the monitoring campaign.

\item The number and nature of monitored appliances, just like the number of observed houses essentially restrain the final usage of databases. In particular, a high number of houses and appliances is required for statistical inspections. In this case, UMSM, Dataport, OCTES, TraceBase, REFIT and HES are the most suitable databases.  Plus, some datasets supervise various houses through multiple time intervals leading to difficulty and even impracticality while comparing between different homes. More than that, the setting under which domestic equipments are employed throughout the day is a basic operator for analyzing the complexity of the usage. This way, experimental campaign should be conducted in real conditions such as households, laboratories, or offices as opposed to simulated environments. 

\item Some databases collect short-term energy consumption and only deliver records of real power, this is the case of COOLL, PSD, ACS-F1 and BLUED. Eventually, seasonal energy usage attitude can not be captured for short-term periods. In this aspect, making use of these databases to track power consumption behavior of end-users is not suitable.

\item A number of databases, among them UMSM, IAWE , ACS-F1, AMPds1, RAE and DRED have furnished a set of electric parameters, including I, V, P, Q, E, f and $\phi$. Additionally to these records, other conditions such as T, O and L are also reported in QUD and REFIT datasets. The latter provides also analytic information notably related to the monitored electrical appliances and integrates statistics about daily activities in dwelling and residential environments, as well. This endows better a interpretive depth comparable to identical repositories (REDD, BLUED, GREEND). 

\item Most of the studied datasets did not capture the exogenous conditions, such as the weather temperature, humidity, which can affect effectively the energy consumption. However, while the REFIT dataset has identical properties to OCTES and ACS-F1 datasets, it is also different because it adds other environmental data including the temperature, light, and motion patterns. In addition, household reviews are also reported, which include the surface, age, heating system, insulation, nature of buildings along with other data specifying the number of individuals, job quality and age. In this context, quantitative statistics gleaned from the reviews with occupants' statistics offer more possibilities to researchers to study the influence of other parameters on energy consumption.

\item There is a lack of available publicly annotated power consumption datasets to train/learn anomaly detection algorithms, in which power consumption variables are clearly labeled as normal or anomalous. Specifically, all the investigated datatsets in this framework except QUD do not encompass labels that identify normal or abnormal consumption, and thereby they can only be used to train unsupervised anomaly detection algorithms because they do not require annotated datasets.

\item Privacy and security concerns have not seriously been considered in most of the existing datasets. This is due to the fact that conventional meters required to be
physically accessed and they registered power consumption for longer time periods (i.e. the real-time monitoring was not considered).

\end{itemize}

\subsection{Qatar university dataset (QUD)}
Using the pros and cons of the state-of-the-art datasets presented in the previous section, a measurement campaign has been conducted in the Qatar university energy lab to glean QUD repository. Specifically, in order to compensate the undersupply of appliance-level datasets dedicated for energy efficiency and anomaly detection in power consumption, a real-time micro-moment laboratory has been developed to gather accurate power usage footprints. Put simply, QUD is a set of consumption records from different installed electrical devices (e.g. air conditioner, heating system, desktop and light lamps) in addition to contextual data; including humidity, temperature, room occupancy and ambient light intensity. To the best of our knowledge, QUD is the first dataset in the Middle East, in which consumption data are collected at an ordinarily 240V voltage. This dataset have multiple usage scenarios such as detecting consumption abnormalities, testing recommender systems and assessing innovative visualization tools. Moreover, it is worth noting that QUD is among the first annotated repositories dedicated for anomaly detection in power consumption.

Therefore, the time-series data representing power consumption footprints for two appliances are registered along with corresponding cubicle occupancy, indoor temperature, indoor humidity, and luminosity. In order to label QUD consumption observations, the micro-moment paradigm is used which helps in identifying the moments of good or anomalous usage. Specifically, the micro-moments are deployed to come up with accurate statistics about consumers \cite{Alsalemi2019is,ALSALEMI2019ae}. Using this dataset, the power consumption observations are labeled via the use of five micro-moment classes  according to a set of standards out of the yielded appliance. These five micro moments are defined as; \enquote{good usage}, \enquote{turn on}, \enquote{turn off}, \enquote{excessive power consumption}, and \enquote{consumption when outside}. The last two micro-moments represent anomalous consumption behaviors that are leading to much wasted energy. Table \ref{micro-labels} describes the micro-moment classes and labels used in QUD (QUD can be accessed through \url{http://em3.i-know.org/datasets/}). In addition, it is worthy to mention that the micro-moment \enquote{consumption when outside} is limited to a set of appliances, such as air conditioners, televisions, light lamps, desktops/laptops, and fans, in which the end-user should be present during their operation to not be considered as an anomalous consumption \cite{HimeurCOGN2020}.

\begin{table} [t!]
\caption{Micro-moments assumption and labeling}
\label{micro-labels}
\begin{center}

\begin{tabular}{l|l|l}
\hline
\textbf{Micro-moment} & \textbf{Label} & \textbf{Description} \\ \hline
\multicolumn{1}{l|}{Good usage} & 0 & Non-excessive usage  \\ \hline
\multicolumn{1}{l|}{Turn on } & 1 & Switching on a device \\ \hline
\multicolumn{1}{l|}{Turn off } & 2 & Switching off a device \\ \hline
\multicolumn{1}{l|}{Excessive consumption} & 3 & Consumption > 95\% of device's maximum active power consumption level \\ \hline
\multicolumn{1}{l|}{Consumption when outside} & 4 & Device consumption without the presence of the end-user \\ \hline
\end{tabular}

\end{center}
\end{table}

\section{Future directions} \label{sec4}
\begin{figure} [t!]
\begin{center}
\includegraphics[width=13cm,height=4.8cm]{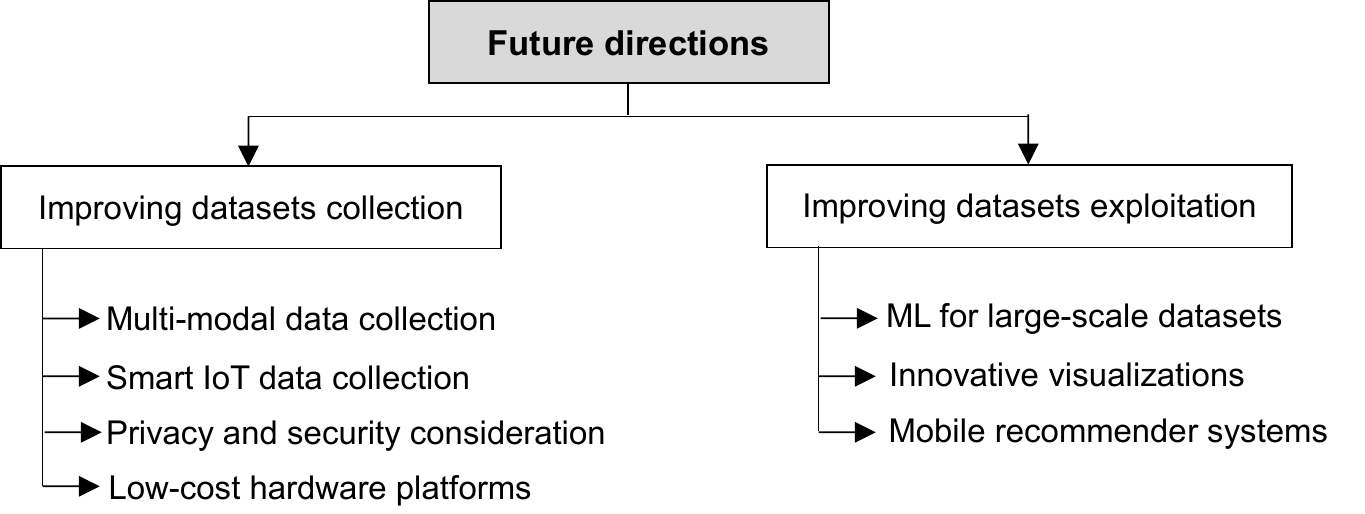}
\end{center}
\caption{Future directions to improve both datasets collection and exploitation.}
\label{Directions}
\end{figure}

After analyzing, comparing and capturing pros and cons of existing datasets, a set of important orientations that can improve data collection and enrich datasets' content are identified. In addition, other directions to improve datasets exploitation are described as well. Figure \ref{Directions} summarizes the future directions that are identified to improve both datasets collection and exploitation.

\subsection{Improving the dataset collection} 
In order to develop powerful energy efficiency systems, it is of paramount importance to improve dataset collection procedures and hence enhance the content of collected data. In this respect, the following recommendations and directions can be establish: 
 
\subsubsection{Multi-modal data collection}
\begin{figure}[t!]
\begin{center}
\includegraphics[width=16cm,height=4.6cm]{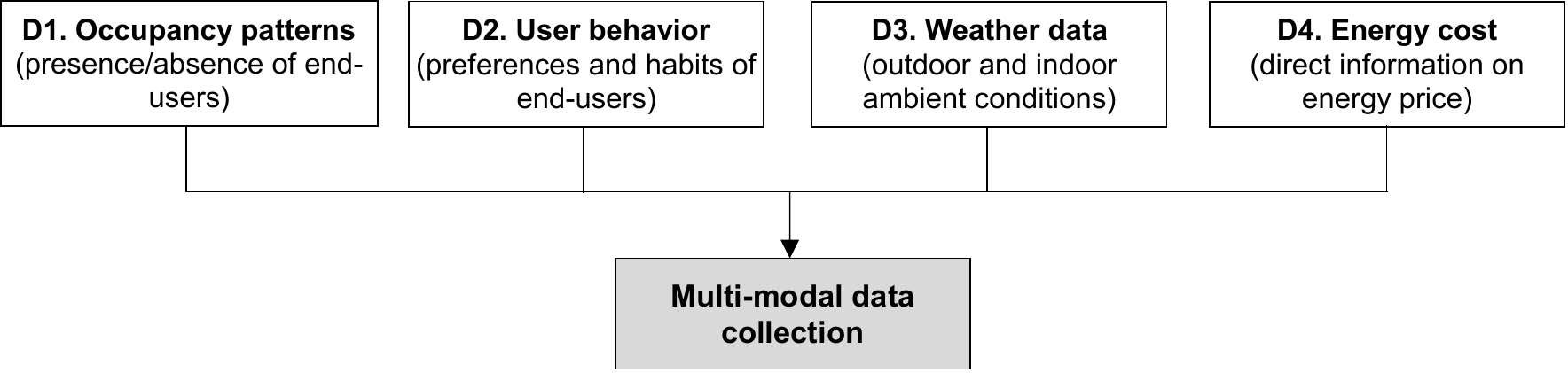}
\end{center}
\caption{Principal factors impacting the power consumption in buildings and contributing in the multi-modal data collection.}
\label{PowerConsume}
\end{figure}

Multi-modal data collection means merely collecting more than one type of data to accomplish an efficient energy saving task or other related applications. Specifiably, power consumption in buildings depends on multiple factors, which should be gleaned together power consumption footprints in order to design comprehensive datasets \cite{Fotopoulou8016276,Himeur2020INFFUS}. Figure \ref{PowerConsume} summarizes the principal parameters impacting the power consumption in buildings and contributing in the multi-modal data collection.  

\vskip2mm
\noindent \textbf{D1. Occupancy patterns:} 
Domestic residents utilize more energy when they're occupied. Even this may appear glaringly evident, collecting occupancy data is a serious matter that must be inspected when searching for wasteful energy aspects in households. Specifically, we ensure that these structures consume less power when unoccupied. Individuals in households influence power consumption for the most part via lighting, cooling, heating and other plug loads. Analyzing power consumption for the duration of the day demonstrates an immediate relationship amongst occupancy and power usage. For the moments when individuals are in a household, different rooms are conditioned or heated to an agreeable temperature. Of course, normal day-by-day activities need also power usage. The effect of individuals utilizing energy in a household is the reason we underscore the relevance of individuals turning off unused appliances or other devices in unoccupied rooms \cite{CAPOZZOLI2017SE,KANG2018}. In this regard, the use of occupancy sensors is highly recommended in households or other buildings such as offices, laboratories or campus buildings to sense when someone is present or not and then turn off the appliances accordingly. By this way, a lot of energy can be preserved when the absence of individuals is confirmed.

\vskip2mm

\noindent \textbf{D2. User behavior:} 
Comprehending and improving individual power consumption behavior is among the successful approaches to reduce energy demand and encourage energy preserving. In fact, user behavior can be responsible of about 20--50\% of the consumption level \cite{DELZENDEH2017,GE2019}. Therefore, collecting and inserting data about end-users' behaviors in the energy efficiency model can significantly decrease wasted energy. This can be done through gathering information related to their preferences and habits \cite{DING2020109939,JIANG2020106410,su10030737}.


\vskip2mm

\noindent \textbf{D3. Weather data:} 
Relation between weather circumstances and power consumption has been proved in several works \cite{KOCI2019,LIU2017,GENG2018}. As a matter of exemplification, peak energy demand during heat waves is widely seen in so many hot countries. For that reason, gathering weather data is regarded as crucial while investigating user behavior. Over and above, existing and newly built households will certainly undergo the impact of climate change. Accordingly, collection and measurement of new energy consumption databases of these houses ought to consider weather patterns that integrate certain repercussions of climate change, rather than only considering historical climate information \cite{FARAH2019,LUPATO2019,ERBA2017}.

\vskip2mm

\noindent \textbf{D4. Energy cost:} 
Estimating the cost of household power consumption and providing this data to end-users can motivate them improving their behavior \cite{ANTONIETTI2019}. Forecasting the user's electricity bill and integrating energy price signals in energy efficiency applications can effectively increase power saving \cite{SATCHWELL2018,ERYILMAZ2018,FIKRU2019} since it helps the consumer to cognitively bridge the gap between consumption and cost. Moreover, collecting energy price profiles at the appliance level makes it unambiguous for the user which appliance raises more the cost. As a result, the consumer can relatively behave in order to reduce wasted energy.

\subsubsection{Smart IoT data collection}
Conventional meters are not able to gather the type of granular and device-level data, however, this becomes possible today with smart meters. In this line, in order to achieve target requirements in relation to data accuracy and further supporting real-time data collection and analysis, deploying smart meters and Internet of things (IoT) sensors to ensure a smart IoT data collection strategy is of paramount importance \cite{Motlagh13020494}. This helps in optimizing the communication, storage and computing resources.

\subsubsection{Low-cost hardware platforms}
In order to reduce the cost of datasets collection, the use of hardware platforms, enabling more cost-effective and powerful alternatives to process and transmit collected data is a high priority, such as the Raspberry PI 4 (RPI4) model B \cite{RPI4}, ODROID-XU4 \cite{odroid-xu4} and Jetson TX1 \cite{JetsonTX1}. Those platforms can monitor energy consumption data along with collecting other essential contextual information, which ultimately results in a larger pool of data.

\subsubsection{Privacy and security consideration}
To preserve the end-users' privacy, power consumption footprints and end-users personal information should be protected. Personal data related to end-user specific power consumption patterns can be exploited to identify and supervise behavior patterns inside buildings (households or public structures). 
This is possible since electrical devices e.g. the microwave, air condition, washing machine, dishwasher, etc. can be detected and recognized from their power consumption fingerprints \cite{HIMEUR2020114877}. Therefore, personal data related to consumption signatures may be deployed to carry out real-time surveillance of end-users. 
In this regard, the data collection process must encourage producing challenging datasets and make power consumption statistics available to end-users and energy providers while respecting end-users' personal privacy and security. To that end, adopting robust techniques to remove personal information is a must, including encryption, steganography and aggregation.

\subsection{Improving the dataset exploitation}
Almost energy efficiency systems are built and validated using energy consumption datasets, which make them very important. Further, with the increasing amount of data collected in each database, the need for challenging solutions that can extract comprehensive information is becoming inevitable \cite{Alsalemi9112672}. In this section we present three main directions, which can be investigated to ameliorate energy saving initiatives. It is worthy to mention that although the following directions are from a consumer's perspective, however, they are valuable for both consumers and energy providers. Specifically, they are generally developed by the latter and deployed to the benefit of end-users to help them in optimizing their energy usage. In addition, as discussed in Section \ref{sec1}, consumers are responsible for wasting more than 20\% of the total energy consumed in buildings \cite{White2019,URGEVORSATZ201585,ALMARRI20173464}.

\subsubsection{Mobile recommender systems}
Lastly, it is noticed that mobile smart devices are becoming an indispensable part of our daily life. Unlike earlier mobile phones that provide limited functionality, smart phones can do a variety of very useful jobs. With the widespread usage of smartphones and the fast growing of the internet and network facilities, a massive amount of data is produced. Consequently, modern societies have started the age of Big Data through successfully discovering users' possible demand and preferences. This has raised the necessity for data scientists and energy management stakeholders to conduct studies on mobile recommender systems for controlling users' energy efficiency \cite{REHABC2020}. 
				
Recommendation systems are commonly deployed to polish the use of smartphones and to assist in dealing with the large amount of data through establishing appropriate advices using recommendation schemes and contextual information. In this regards, the role of recommender systems will be essential to promote energy efficiency and help end-users understanding and improving their consumption footprints \cite{Ricci2010}. More specifically, every particular recommender application is generally elaborated with an explicit context in mind with the aim of solving in some sense the data overloading issue due to the large-scale datasets of power consumption. The effectiveness of a mobile recommender system has been demonstrated through real-use applications in academic buildings \cite{REHABC2020}, in which a context-aware based recommender app is developed to help in supporting end-users to transform their energy consumption habits.

Furthermore, the architecture of recommendation systems that is usually based on interactional models, graphic user interfaces and recommendation engines makes them productive and useful to deal with energy efficiency applications \cite{Starke2017acm}. To summarize, the use of mobile recommender systems is recommended to improve power consumption datasets exploitation via:

\begin{itemize}
\item Developing explainable recommender systems can be very supportive to improve data exploitation with a view to replacing inefficient energy habits with efficient ones. An explainable recommender system aims at providing end-users with tailored recommendations, followed by explanations about them \cite{zhang2020explainable}. Explanations refer to the motivations behind the recommendation or to the benefits from providing the recommended action or advice. They can enhance the persuasiveness of the system, end-users' understanding and satisfaction and provide an immediate reward to them. 
\item Developing intelligent mobile home monitoring systems using collected data to provide information and monitoring options to the end-user to help him control its load usage, visualize consumption statistics and compare them to those of other users, and further predict the overall charge of monthly bills \cite{Neurio2019}. 
\end{itemize}

\subsubsection{Visualization for understanding user behavior} \label{sec-visual}
\begin{figure}[t!]
\begin{center}
\includegraphics[width=14cm,height=6.5cm]{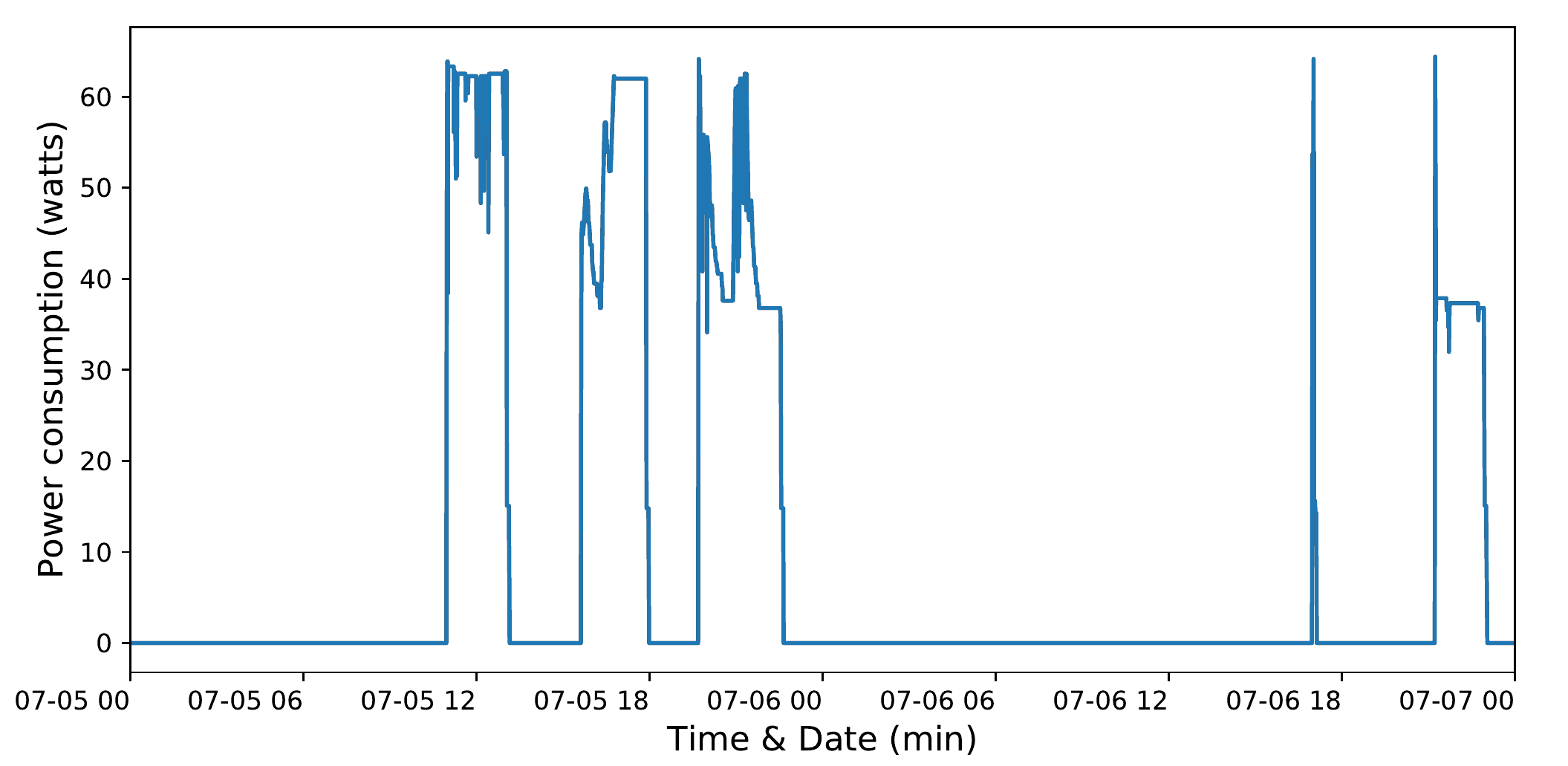}
\includegraphics[width=14cm,height=6.5cm]{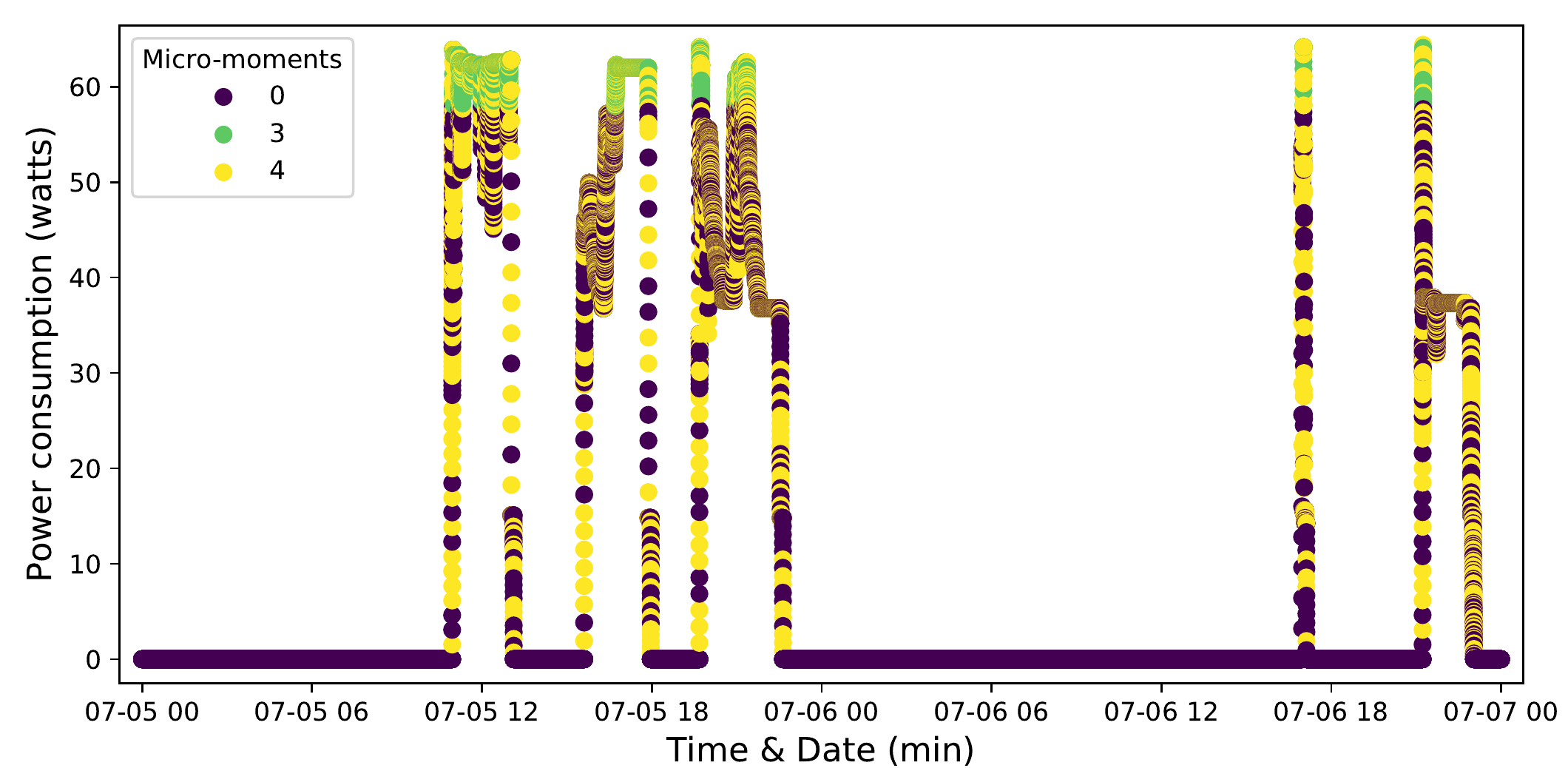}
\end{center}
\caption{Time-series power consumption of a television and its micro-moments scatter plot from DRED: top) time-series power consumption, and bottom) micro-moments scatter plot at a sampling rate of 3 min.}
\label{scatter_micro}
\end{figure}

Visualization is seen to be the most effective way to assimilate increasingly large datasets with the aim of interactively and perfectly conveying insights to end-users, consumers, and stakeholders in general. Recent tools, methods, and softwares leveraged for visualization of energy consumption require further improvements to remain more important in a planet with larger low-carbon emissions. Moreover, they are required to sensitize energy-consuming behavior in an approachable and stimulating way. In this context, we present in this section an example of a novel visualization approach based on micro-moments analysis. Figure \ref{scatter_micro} displays a time-series energy consumption of a television and its micro-moments scatter plot at sampling intervals of 3 min, recorded in DRED dataset. This novel visualization strategy is presented as an example, in which energy usage micro-moments of 2 days are captured and plotted. It is used to display good usage (class:0), excessive usage (class: 3) and consumption while outside (class: 4).
Users can seamlessly get the plots at different sampling rate starting from the milliseconds. 

As it can be deduced, tracing micro-moments through time patterns facilitates identifying moments of abnormal consumption and then makes it easy to establish precise guidelines helping to reduce energy waste. Moreover, this helps end-users understanding their consumption footprints, increasing their awareness, and hence triggering them to improve their behavior through the use of tailored recommendations. In addition anomalous consumption behaviors can be identified when an adequate visualization tool is adopted, e.g. the micro-moments visualization, and hence end-users can improve their behaviors based on the detected anomaly. Moreover, it is worth noting that the use of the micro-moments paradigm to detect anomalous consumption can be enlarged to identify other kinds of anomalies, e.g. detecting abnormal consumption of an air conditioner while doors/windows are open via considering other information sources. Therefore, end-users will be provided with the appropriate notifications and advices, i.e. close doors/windows to reduce wasted energy.

In addition, a set of valuable recommendations and future directions towards designing effective visualizations aiming to increase end-users energy awareness is summarized as follows: 
\begin{itemize}
\item Visualizations need to catch the attention of their users via using bright colors, contrasts and varied views, where somethings are changing constantly. More importantly, they should implement colors that are legible for people with color vision deficiencies \cite{TORABIMOGHADAM201919}. 
\item Visualizations also need to employ comparisons between different time period consumptions to stimulate energy users' reflection and learning. Individuals are highly motivated in preserving power usage through making comparison of their current consumption to their own previous consumptions.
\item Fragile long time-series aggregated systems for visualizing energy consumption are better to be avoided. By nature, the human being's mind can not always memorize all what they do in each second or which appliance do they use more in their daily routine. Hence, it was concluded by many studies that: 

\noindent \textbf{a.} Collecting appliance-level consumption data is highly needed for consumers to cognitively decode their consumption data and to finally make a decision towards changing their behavior \cite{Melanie2017}. 

\noindent \textbf{b.} Instantaneous short-time intervals are the best way to aid the consumer to easily figure out at what instance their energy cost increased \cite{Spence2018}. This makes it possible for the users to ameliorate their behavior and to notice the direct effect from it. 

\item 3D visualizations of households with real-time energy consumption statistics can impart better contextual understanding to homeowners and help them control practices shaping their consumption. In relation, it is also crucial to support that by exploring the deployment of other emerging technologies, including virtual and augmented reality, interactive visualizations and wall sized presentations. 

\end{itemize}

\subsubsection{ML algorithms of large-scale datasets}
Another important direction that can improve the quality and exploitation of power consumption datasets relies on deploying ML algorithms, which can help significantly in reducing energy consumption and thereby decreasing energy costs and carbon dioxide emissions \cite{LI2019,ZEKICSUSAC2020102074}. Therefore, boosting novel progresses and challenges with regard to ML models is of paramount importance. In this context, several directions could be identified in which ML play a major role when shifting to more sustainable end energy efficiency environment, among them:

\begin{itemize}
\item The use of generative models, such as generative adversarial networks (GANs), which can tremendously improve the quality of collected data via completing incomplete power consumption signals (due to data loss occurred during the collection step) and hence leading to a better exploitation of the produced datasets in different applications\cite{Fekri2019}.

\item The use of deep learning models to identify consumption anomalies via classifying the micro-moment classes of end-users' power consumption \cite{ALSALEMI2019ae}. These algorithms assist in analyzing consumption footprints to detect abnormal consumption related to \enquote{excessive consumption} and \enquote{consumption when outside}. After defining the micro-moments described in section \ref{sec-visual}, DNN and other ML algorithms with different configuration parameters could be deployed, including logistic regression, linear discriminant analysis (LDA), support vector machine (SVM), K-nearest neighbors (KNN), decision Tree (DT) and ensemble bagged tree (EBT. The selection of the appropriate ML model is mainly based on ensuring the best compromise between the identification performance and computational complexity. Therefore, this results in a better exploitation of the collected datasets for identifying abnormal power consumption.

\end{itemize}

\section{Conclusion} \label{sec5}
{ 
Existing building power consumption datasets have been reviewed together with their merits and drawbacks within the lines of this work. Comprehensive comparisons between these datasets have been conducted in terms of different factors and data collection platforms that categorize each of these. Based on the fruitful analysis gleaned from these comparisons, a novel annotated dataset has been presented, namely QUD, which can be very useful for power consumption anomaly detection since it includes labels of good and anomalous usage. Moreover, we came with recommendations and future directives for improving the quality and the content of future power consumption datasets. Thus, a set of relevant orientations have been identified as follows:
\begin{itemize}
\item Adopting a multi-modal data collection, which is based on gathering various data sources because the energy consumption is affected by different factors. Therefore, data should be gathered from several geographical positions with regard to ambient conditions, atmospheric and environmental resources, occupation patterns, user preferences and energy cost.
\item Adopting smart IoT data collection strategies to collect data from various IoT sensors at a low sampling rate, which leads therefore to developing real-time and scalable energy saving systems.
\item Collecting more annotated anomaly detection datasets in order to encourage testing and developing anomaly detection algorithms. Overall, detecting anomalous power consumption plays a major role in reducing wasted energy.
\item Producing comprehensive databases by reference to the level of consumption (especially at the appliance-level) and the duration of the collection campaign (i.e. collecting data for the entire seasonal periods of the year) while respecting end-users personal privacy and security
\item Formulating protocols and standards to characterize building power consumptions datasets that can help make unified metadata strategies and terminologies, facilitating the comparisons, and rigorously understanding the state-of-the-art.
\end{itemize}

In addition, genuine initiatives to improve datasets exploitation and utilization have been identified in this framework. Consequently, a novel visualization strategy has been presented based on the micro-moments analysis, which enables people to comprehend their own power usage footprints, and accordingly interpret their electricity consuming behavior. Moreover, it helps them easily getting statistics on their actual power consumption. Moreover, another example of using ML algorithms has been introduced to classify power consumption micro-moments and detect anomalous usage, such as \enquote{excessive consumption} or \enquote{consumption while outside}. These two behaviors are responsible for wasting a large amount of energy. Consequently, it will be part of our future work to improve datasets exploitation and help promoting energy efficiency through: 
\begin{itemize}
\item The use of novel ML algorithms, including deep learning and generative adversial networks (GAN), which can effectively deal with imbalanced and large-scale datasets.
\item The development of innovative visualization tools that can cognitively improve end-users comprehension of their consumption behavior. Therefore, interactive visualization tools will be integrated into smart power consumption dashboards that permit individuals to engage with, apprehend their energy usage and translate them into positive actions throughout their everyday life.
\item The deployment of explainable recommender systems in order to trigger action recommendations at the correct moment, especially if appropriate hardware materials are used to collect and analyze data in a real-time manner. 
\end{itemize}

\section*{Acknowledgements}\label{acknowledgements}
 This paper was made possible by National Priorities Research Program (NPRP) grant No. 10-0130-170288 from the Qatar National Research Fund (a member of Qatar Foundation). The statements made herein are solely the responsibility of the authors.

\section*{Appendix }
Abbreviation description of the power consumption datasets along with a description of the nomenclatures considered in this paper are summarized in Table \ref{AbbDescription}.

\begin{table}[htbp]
\caption{List of abbreviations and nomenclatures used in this paper.}
\label{AbbDescription}
\begin{center}

\begin{tabular}{lc|cc}
\hline
{\small Acronym} & {\small Description} & {\small Nomenclature} & {\small %
Description} \\ \hline
{\small UMSM} & \multicolumn{1}{l|}{\small UMass Smart* Microgrid} & 
\multicolumn{1}{|l}{\small I} & \multicolumn{1}{l}{\small current} \\ 
{\small IHEPCDS} & \multicolumn{1}{l|}{\small Individual household
electric power consumption data set} & \multicolumn{1}{|l}{\small V} & 
\multicolumn{1}{l}{\small voltage} \\ 
{\small IHEPCD} & \multicolumn{1}{l|}{\small Household electricity survey} & 
\multicolumn{1}{|l}{\small P} & \multicolumn{1}{l}{\small active power} \\ 
{\small REFIT} & \multicolumn{1}{l|}{\small Personalized retrofit decision
support tools for UK homes} & \multicolumn{1}{|l}{\small Q} & 
\multicolumn{1}{l}{\small reactive power} \\ 
{\small ORBEET} & \multicolumn{1}{l|}{\small Opportunities for community
groups through energy storage} & \multicolumn{1}{|l}{\small S} & 
\multicolumn{1}{l}{\small apparent power} \\ 
{\small AMPds1} & \multicolumn{1}{l|}{\small Almanac of minutely power
dataset version 1} & \multicolumn{1}{|l}{\small Np} & \multicolumn{1}{l}%
{\small normalized power} \\ 
{\small ECB} & \multicolumn{1}{l|}{\small Electricity consumption benchmarks} & \multicolumn{1}{|l}{\small E} & \multicolumn{1}{l}{\small energy}
\\ 
{\small PSD} & \multicolumn{1}{l|}{\small Pecan street dataset} & 
\multicolumn{1}{|l}{\small f} & \multicolumn{1}{l}{\small frequency} \\ 
{\small MEULPv.1} & \multicolumn{1}{l|}{\small Measured end-use electric
Load profiles version 1} & \multicolumn{1}{|l}{${\small \phi }$} & 
\multicolumn{1}{l}{\small phase angle} \\ 
{\small RAE} & \multicolumn{1}{l|}{\small Rainforest automation energy} & 
\multicolumn{1}{|l}{\small pf} & \multicolumn{1}{l}{\small power factor} \\ 
{\small GREEND} & \multicolumn{1}{l|}{\small Green dataset} & 
\multicolumn{1}{|l}{\small EC} & \multicolumn{1}{l}{\small energy cost} \\ 
{\small ECO} & \multicolumn{1}{l|}{\small Electricity consumption and
occupancy} & \multicolumn{1}{|l}{\small Wt} & \multicolumn{1}{l}{\small %
weather} \\ 
{\small RDED} & \multicolumn{1}{l|}{\small Dutch residential energy dataset}
& \multicolumn{1}{|l}{\small T} & \multicolumn{1}{l}{\small temperature} \\ 
{\small IWAE} & \multicolumn{1}{l|}{\small Indian dataset for ambient water
and energy} & \multicolumn{1}{|l}{\small H} & \multicolumn{1}{l}{\small %
humidity} \\ 
{\small DISEC} & \multicolumn{1}{l|}{\small Dataset on information
strategies for energy conservation} & \multicolumn{1}{|l}{\small O} & 
\multicolumn{1}{l}{\small occupancy} \\ 
{\small CRHLP} & \multicolumn{1}{l|}{\small Commercial and residential
hourly load profiles dataset} & \multicolumn{1}{|l}{\small L} & 
\multicolumn{1}{l}{\small light level} \\ 
{\small HUE} & \multicolumn{1}{l|}{\small Hourly usage energy dataset} &  & 
\\ 
{\small UK-DALE} & \multicolumn{1}{l|}{\small UK domestic appliance-level
electricity} &  &  \\ 
{\small REDD} & \multicolumn{1}{l|}{\small Reference energy disaggregation
dataset} &  &  \\ 
{\small BLUED} & \multicolumn{1}{l|}{\small Building-level fully labeled
electricity disaggregation dataset} &  &  \\ 
{\small BLOND} & \multicolumn{1}{l|}{\small Building-level office
environment dataset} &  &  \\ 
{\small PLAID} & \multicolumn{1}{l|}{\small Plug Load Appliance
Identification Dataset} &  &  \\ 
{\small ACS-F1} & \multicolumn{1}{l|}{\small Appliance consumption signatures version 1}
&  &  \\ 
{\small ENERTALK} & \multicolumn{1}{l|}{\small Energy talks} &  &  \\ 
{\small DRED} & \multicolumn{1}{l|}{\small Dutch residential energy dataset}
&  &  \\ 
{\small REFIT} & \multicolumn{1}{l|}{\small Retrofit decision support tools
for UK homes} &  &  \\ 
{\small OCTES} & \multicolumn{1}{l|}{\small Opportunities for community
groups through energy storage} &  &  \\ 
{\small COOLL} & \multicolumn{1}{l|}{\small Controlled On/Off Loads Library}
&  &  \\ 
{\small MEULPv.2} & \multicolumn{1}{l|}{\small Measured end-use electric
Load profiles version 2} &  &  \\ 
{\small BERDS} & \multicolumn{1}{l|}{\small Berkeley energy disaggregation
dataset} &  &  \\ 
{\small CRHLP} & \multicolumn{1}{l|}{\small Commercial and residential
hourly load profiles dataset} &  &  \\ 
{\small IoT} & \multicolumn{1}{l|}{\small Internet of things} &  &  \\ 
{\small LDA} & \multicolumn{1}{l|}{\small Linear discriminant analysis} &  & 
\\ 
{\small SVM} & \multicolumn{1}{l|}{\small Support vector machine} &  &  \\ 
{\small KNN} & \multicolumn{1}{l|}{\small K-nearest neighbors } &  &  \\ 
{\small DT} & \multicolumn{1}{l|}{\small Decision tree} &  &  \\ 
{\small EBT} & \multicolumn{1}{l|}{\small Ensemble bagged tree} &  &  \\ 
{\small DNN} & \multicolumn{1}{l|}{\small Deep neural networks} &  &  \\ 
{\small MLP} & \multicolumn{1}{l|}{\small Multilayer perceptron} &  &  \\ 
{\small LR} & \multicolumn{1}{l|}{\small Logistic regression} &  &  \\ 
{\small NILM} & \multicolumn{1}{l|}{\small Non-intrusive load monitoring} & &  \\ 
{\small JSON} & \multicolumn{1}{l|}{\small JavaScript Object Notation} &  & \\ 
{\small SQL} & \multicolumn{1}{l|}{\small Structured query language} &  & \\ 
{\small ML} & \multicolumn{1}{l|}{\small Machine learning} &  & \\ 
{\small GANs} & \multicolumn{1}{l|}{\small Generative adversarial networks} &  & \\ 

{\small (EM)$^{3}$} & \multicolumn{1}{l|}{\small Endorsing energy efficiency with
micro-moments} &  &  \\ \hline
\end{tabular}

\end{center}
\end{table}

\end{document}